\documentclass[pdflatex,sn-standardnature]{sn-jnl}

\usepackage{amsmath}
\usepackage{lineno}
\usepackage{natbib}
\usepackage{mathtools}

\jyear{2023}%

\theoremstyle{thmstyleone}%
\theoremstyle{thmstyletwo}%

\theoremstyle{thmstylethree}%

\begin{document}

\title[Statistics of modal condensation in nonlinear multimode fibers]{Statistics of modal condensation in nonlinear multimode fibers}


\author*[1]{\fnm{Mario} \sur{Zitelli}}\email{mario.zitelli@uniroma1.it}
\author[1]{\fnm{Fabio} \sur{Mangini}}\email{fabio.mangini@uniroma1.it}
\author[1]{\fnm{Stefan} \sur{Wabnitz}}\email{stefan.wabnitz@uniroma1.it}




\affil*[1]{\orgdiv{Department of Information Engineering, Electronics and Telecommunications}, \orgname{Università degli Studi di Roma Sapienza}, \orgaddress{\street{Via Eudossiana 18}, \city{Rome}, \postcode{00184}, \state{RM}, \country{Italy}}}




\abstract{Optical pulses propagating in multimode optical fibers are affected by linear disorder and nonlinearity, and experience chaotic exchange of power among modes. On the other hand, complex systems can attain steady states characterized by energy condensation into single as well multiple sub-systems. In this work, we study beam propagation in multimode optical fibers in the presence of linear random mode coupling and Kerr nonlinearity; both effects lead to a mode power redistribution at the fiber output. We use a new 3D mode decomposition method to obtain, with unprecedented accuracy, measurements of the modal distribution from long spans of graded-index fiber; we perform numerical simulations using a new model for the linear disorder; we introduce a weighted Bose-Einstein law and show that it is suitable for describing steady-state modal power distributions both in the linear and nonlinear regimes. We show that, at power levels intermediate between the linear and the soliton regimes, energy condensation is attained locally by the second, third and fourth modal groups, before global condensation to the fundamental mode is reached in the soliton regime. Our results extend the thermodynamic approach to multimode fibers to unexplored optical states, which acquire the characteristics of optical glass.}




\maketitle

\section{Introduction}\label{sec:sec0}

Multimode (MM) optical fibers \cite{6767859, https://doi.org/10.1002/j.1538-7305.1972.tb01911.x,Hasegawa:80} have regained considerable interest over the last decade, motivated by the potential for increasing the transmission capacity using the technique of spatial-division multiplexing (SDM) \cite{Richardson-NatPhot-2013-94-2013,Winzer:18}, and the possibility of up-scaling the pulse energy delivered from fiber lasers \cite{wright2017spatiotemporal}.

In recent years, a thermodynamic interpretation of beam propagation in multimode optical fibers and systems has been formulated \cite{Wu2019-nn,Fusaro_PhysRevLett.122.123902}. In the specific example of a graded-index (GRIN) MM fiber, photons are divided into indistinguishable energy packets: statistical mechanics permits to estimate the equilibrium distribution of these packets among degenerate groups of modes, with the same propagation constant $\beta_j$ (m$^{-1}$); such equilibrium distribution corresponds to extrema of the entropy $S$ of the mode population.

\begin{figure}[h]
\includegraphics[width=0.95\textwidth]{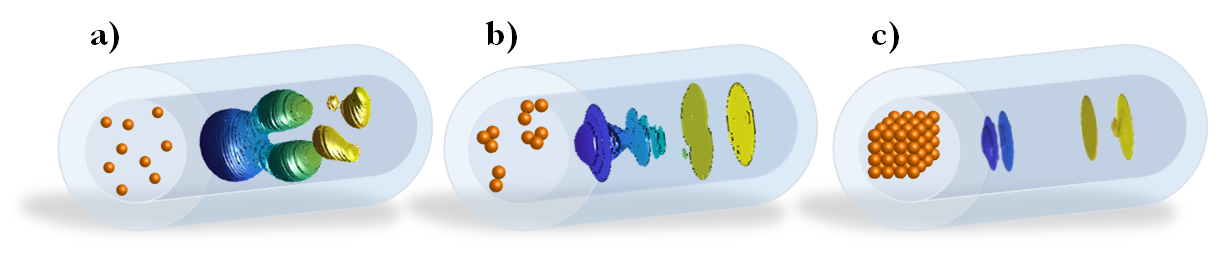}	\centering	
\caption{Optical bullets forming after long spans of GRIN fiber in a) linear, b) nonlinear and c) soliton regimes. A thermodynamic representation of the three states is artistically represented.}
\label{fig:Fig1}
\end{figure}

On the other hand, it is well known that optical pulses propagating in a MM fiber in the linear regime are affected by random-mode coupling (RMC) \cite{Gloge:6774107,Marcuse1973LossesAI,Ho:14}, which is induced by fiber imperfections such as micro and macro-bending. Pulses in modes within degenerate (i.e., with the same or nearly the same propagation constant) groups separate in time from pulses in other groups, as a consequence of inter-modal dispersion. 3D bullets carried by each group broaden in time because of intra-modal or chromatic dispersion (Fig.\ref{fig:Fig1}a). In the linear regime, RMC effects can be modelled by power-flow equations \cite{Savovi2019PowerFI,Gloge:6774107,Olshansky:75}, which predict a diffusion of energy from intermediate groups into both lower-order and high-order modes. These models lead, after few hundreds meters of propagation, to a steady-state mode power distribution showing a characteristic decrease of power as the mode order grows larger.

At higher pulse energy, in MM fibers with anomalous chromatic dispersion, optical solitons form with a specific pulsewidth $T_{\text{FWHM}}$ depending on the pulse wavelength \cite{Zitelli2021a}, and with energy given by $E_s=1.76\lambda \lvert \beta_2(\lambda) \rvert w_e^2/(n_2T_{\text{FWHM}})$, with $\beta_2(\lambda)$ (s$^2$/m) the chromatic dispersion, $n_2$ m$^2$/W the Kerr nonlinear coefficient, and $w_e$ the modal effective waist.
At telecom wavelength $\lambda=1550$ nm, the soliton pulsewidth is $T_{\text{FWHM}}=120$ fs. Pulses corresponding to the different modal groups get shorter and increase their peak power; Kerr and Raman nonlinearities induce a slow but monotonic transfer of energy from each group to the fundamental mode \cite{Zitelli2021}, leading to global condensation in the ground state. When pulses separate in time, the mode coupling process is eventually continued by the interplay of RMC and IM-FWM. After hundreds of meters of propagation, a train of fundamental soliton bullets remains, which experience Raman soliton self-frequency shift \cite{Gordon1986,Zitelli_9887813} (Fig.\ref{fig:Fig1}c). The process can be viewed as a new fission mechanism, mediated by the modal dispersion, as it will be illustrated in Sec \ref{subsec:sec4}.


At intermediate energy, 20\% to 80\% of the soliton $E_s$, a quasi-soliton forms, characterized by pulses carrying the individual modal groups which are initially overlapped in time, and reduce their pulsewidth to values approaching the soliton $T_{\text{FWHM}}$. Mode power exchange is dominated by the IM-FWM, while Raman self-frequency shift is not able to affect the pulse wavelength yet; a conservative wave propagation system can still be assumed. In this regime, local condensation of energy among the lower-order groups is observed (Fig.\ref{fig:Fig1}b); the modes assume a characteristic distribution, which cannot be traced back to what has been observed in the linear or soliton regimes. The local condensation observed in MM shows similarities to those in disordered lasers \cite{PhysRevLett.101.143901} and Bose-Einstein condensation \cite{Klaers_2011}.

In the following sections, a new weighted Bose-Einstein (BE) equation will be demonstrated, which corresponds to extrema of the entropy of an optical multimode system (Sec. \ref{sec:sec1}). Experiments will be shown in long spans of graded-index (GRIN) multimode fibers, from linear to soliton regimes (Sec. \ref{sec:sec2}); modal content emerging from the fiber will be analyzed using the weighted BE in order to check the achievement of local or global condensation states. Numerical power-flow simulations will be performed in the linear regime in Sec. \ref{subsec:sec3}, and analyzed using the weighted BE. Finally, numerical simulations using coupled-mode generalized nonlinear Schröedinger equations (GNLSE), including nonlinearity and a new model for RMC (Sec. \ref{subsec:sec4}), will be used for comparison with the experimental data. New steady states will be found to form in the quasi-soliton regime, characterized by local condensates which can be viewed as glassy states  \cite{parisi_urbani_zamponi_2020,PhysRevB.83.134204}; this condition appears as intermediate between the low-energy disordered state, and the high-energy, highly condensed soliton state.

\section{Weighted Bose-Einstein Law}\label{sec:sec1}

We start considering an optical multimode system including $Q$ groups of degenerate modes, distributed over $g_j/2$ modes and 2 polarizations ($g_j$ is the group degeneracy), with $j=1, 2, .., Q$; in the special case of a graded-index multimode fiber (GRIN), it is $g_j=2, 4, 6,.., 2Q$. Let $n_j$ be the population of energy packets into the j-th group over 2 polarizations; packets are not necessarily single photons ( groups of indistinguishable photons can compose a single packet); hence, the value of $n_j$ is of the order of 10$^5$ to 10$^9$ in practical applications. Extremization of the system's entropy $S$, with the proper choice of Lagrange multipliers, brings to the weighted Bose-Einstein distribution for the modal power fractions (see Methods Sec. \ref{subsec:subsec6_1})

\begin{equation}
\lvert f_i \rvert ^2=\frac{2(g_i-1)}{g_i\gamma}\frac{1}{\exp\big(-\frac{\mu'+\epsilon_i}{T}\big)-1}    .
\label{eq:BE}
\end{equation}

In Eq.~\ref{eq:BE}, $\lvert f_i \rvert^2=2 n_j/(\gamma n_0 g_i)$ is the mean modal power fraction over two polarizations, with $\gamma$ given by the total number of energy packet $N=\gamma n_0$ and $n_0$ a reference number of packets (for example the value at the lowest tested power). $i$ is the mode index in the $j$-th group. In the case of a GRIN fiber, there are $2M=Q(Q+1)$ modes and polarizations, and $i=1, 2,..M$; it is $i=1$ for $j=1$, $i=2, 3$ for $j=2$, $i=4, 5, 6$ for $j=3$ and so on. $g_i=[2, 4, 4, 6, 6, 6,.., 2Q]$, $\epsilon_i=\beta_i-\beta_{j=Q}$ are the differential modal eigenvalues with $\beta_i$ (m$^{-1}$) the propagation constants.
Eq.~\ref{eq:BE} was obtained under the reasonable condition $n_0>>\exp{\big(-\frac{\mu'+\epsilon_i}{T}\big)}-1$; it is able to reproduce the modal power distribution both in linear regime, where RMC is mostly responsible for power exchange among modes, and in nonlinear regime, where inter-modal four-wave mixing (IM-FWM) dominates. Chemical potential $\mu'$ (m$^{-1}$), temperature $T$ (m$^{-1}$) and the normalized power $\gamma$ are three degrees of freedom for fitting Eq.~\ref{eq:BE} to the experimental data, with the only constraint for $\gamma$ to scale with the input power, and to respect the conservation law $\sum_{i=1}^{M}\lvert f_i \rvert ^2=1$.

The thermodynamic validity of Eq.~\ref{eq:BE} can be stressed using the state equation in Methods Sec. \ref{subsec:subsec6_1}, Eq.~\ref{eq:StateEquation2}, which can be reformulated as

\begin{equation}
SE=\sum_{j=1}^Q \beta_j j \lvert f_j \rvert^2 +\mu'-\beta_{j=Q} +(2M-Q) \frac{T}{\gamma}=0 ;
\label{eq:StateEquation3}
\end{equation}

an experimental error upon the state equation can be calculated as

\begin{equation}
\epsilon_{SE}=\frac{SE}{\sum_{j=1}^Q \beta_j j \lvert f_j \rvert^2 -\Big(\mu'-\beta_{j=Q} +(2M-Q) \frac{T}{\gamma}\Big)} .
\label{eq:StateEquationError}
\end{equation}

In Eq.~\ref{eq:StateEquation3}, the first term in the sum (the normalized energy) must be constant; $\mu'$ and $T/\gamma$ must scale in order to maintain constant the remaining part of the equation. Not necessarily the parameters $T/\gamma$ and $\mu'$ must remain separately constant.

In short fiber spans at low power, both RMC and IM-FWM are negligible; we can assume equal modal distributions at the input and output of the fiber $\lvert f_j^{(in)} \rvert^2=\lvert f_j^{(out)} \rvert^2$. Since the input modal distribution is unchanged at high power, Eq.~\ref{eq:StateEquation3} can be used, together with $\lvert f_j^{(out)} \rvert^2$ measured at low power, to predict the thermodynamic parameters of the system at high power, when thermalization, or condensation, is achieved \cite{pourbeyram2022direct}. However, in long spans of fiber at low power, RMC causes $\lvert f_j^{(in)} \rvert^2 \ne \lvert f_j^{(out)} \rvert^2$; thermodynamic parameters are found in this case by fitting Eq.~\ref{eq:BE} to the experimental distribution at a given power; Eq.~\ref{eq:StateEquation3} is used to verify the thermodynamic validity of the fit, but not for parameter predictions from low power experiments.

The validity of Eq.~\ref{eq:BE} is also subject to the constraints of negligible variations of the power $P$ and the energy $U$ (Sec. \ref{subsec:subsec6_1}).
Linear losses are negligible, at the telecom wavelength, for fiber length up to few kilometers. 
Raman nonlinearity is responsible for a frequency shift of the pulse spectrum, and for the consequent variation in the packet energy; in soliton propagation, Raman self-frequency shift (RSFS) \cite{Gordon1986} induces a red-shift of the spectrum without distortion; assuming a tolerance of 5\% on the energy change, a red-shift of 80 nm at 1550 nm wavelength can be tolerated in the experiments.
Chromatic dispersion broadens pulses in the linear regime, considerably reducing the pulse peak power $P$; however, in the quasi-soliton and soliton regimes, the propagating train of pulses conserve their pulsewidth, and negligible changes of power can be assumed over large portions of the fiber.
RMC is also responsible for negligible energy variations; let us assume, by way of example, the propagation at 1550 nm in a GRIN fiber, and that RMC is responsible for a complete power transfer from group 10 to the fundamental mode (this is, of course, a worst case); the fractional energy change equals the one of the propagation constants $(\beta_1-\beta_{10})/\beta_{1}=0.0078$. In practical cases, strong RMC induces negligible energy variations, comparable to the linear losses from few meters of fiber. We can assume then, that the presence of strong RMC does not invalidate the thermodynamic approach.  

Experimental modal distributions that are suitably fitted by Eq.~\ref{eq:BE} correspond to extrema of the entropy $S$ denoting the achievement of a steady state, eventually characterized by a local or global condensation; power fluctuations related to local condensation do not invalidate the equation, which is used to fit the power distribution of the modal groups as a whole; in the next section, Eq.~\ref{eq:BE} will be used for this purpose. 


\section{Experiments}\label{sec:sec2}

In order to determine the mode power distribution at the output of GRIN MM fibers, we used the mode-decomposition method introduced in \cite{Zitelli:23}. Details about our experimental setup are provided in Methods Sec.~(\ref{subsec:subsec6_2}). We have progressively varied the input pulse energy $E_{in}$, so that we could explore the whole spatio-temporal propagation regime from the linear to the nonlinear case. We used 250 fs full-width-at-half-maximum (FWHM) pulses at 1400 nm, with 100 KHz repetition rate. The input beam was circularly polarized, and coupled with a waist of 13 $\mu$m and a
10 $\mu$m lateral shift with respect to the fiber axis, in order to enhance the population of higher-order modes (HOM), and minimize power exchange between polarizations. We used commercial OM4 GRIN (Thorlabs GIF50E) fibers, with lengths of 1 m, 830 m and 5 km, respectively, spooled on a support with radius of curvature larger than 8 cm. 



\begin{figure}[t!]
\includegraphics[width=0.7\textwidth]{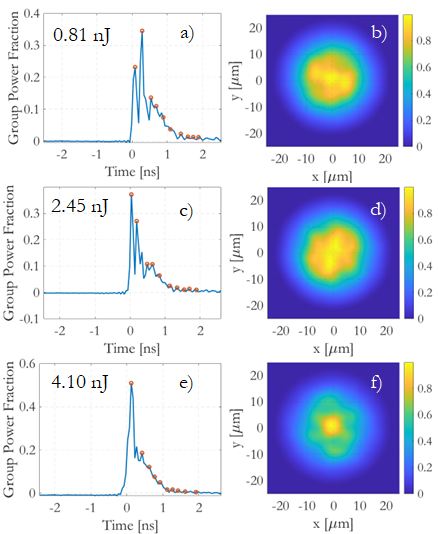}	\centering	
\caption{Normalized group power fractions (left) with their corresponding near-field intensities (right) after 830 m of GRIN fiber, for input pulse energy of a-b) 0.81 nJ, c-d) 2.45 nJ, e-f) 4.10 nJ.}
\label{fig:Fig2}
\end{figure}

Fig.\ref{fig:Fig2} shows the normalized instantaneous output power (left) and the near-field (right) after 830 m of GRIN. Modal dispersion leads to the observed time delay among pulses carried by different groups of degenerate modes. Note that the sub-pulses are equally spaced in time, owing to the equal spacing of the mode propagation constants in GRIN fibers. The pulse carried by the HOMs has the largest delay, hence it appears in the trailing tail of the output waveform. 

Using a fiber which is long enough to temporally separate from different mode groups permits to directly measure the output mode power distribution. This results from the combined action of linear and nonlinear mode coupling, which occurs before pulse splitting. The linear and nonlinear interaction among different modal groups is mostly concentrated in the first portion of the fiber, where pulses overlap in time. Hence, negligible linear losses can be assumed over the interaction length, thus permitting a correct thermodynamic interpretation of the results.

In the linear regime, we estimate that chromatic dispersion (CD) broadens the individual pulses in each group up to a 110 ps duration after 830 m (when considering the nominal -12 ps$^2$/km chromatic dispersion). Modal dispersion produces a time delay of 206 ps between adjacent group pulses.
Optical pulse energy is varied from linear regime (0.1 nJ) to soliton regime (4.5 nJ). Quasi-soliton regime is obtained from 1.0 to 4.1 nJ energy, characterized by trains of short pulses. Raman soliton regime is reached at 4.5 nJ: no Raman delay or spectral shift was observed up to 4.1 nJ; hence, a conservative system is assumed up to 4.1 nJ.

In the linear regime (Fig.\ref{fig:Fig2}a), the output distribution of energy among the mode groups is determined by linear disorder. For $E_{in}=0.81$ nJ, the effective length of the fiber (determined by taking into account both the weak linear loss and the rapid pulse peak power decrease owing to chromatic dispersion-induced temporal broadening) is as short as $L_{eff}=12$ m. This means that, for most of the distance, power exchange among pulses carried in different modes is only determined by RMC. 
 

In Fig.\ref{fig:Fig2}a, c, e, we sample the photodiode traces at points corresponding to the relative group delay of each modal group (see the orange circles), which permits a mode power decomposition as described in \cite{Zitelli:23}. In this approach, and considering the first $Q=10$ modal groups, corresponding to $M=Q(Q+1)/2=55$ modes per polarization, one directly measures the mode group powers $P_j$. The corresponding mean modal power fraction carried by each mode in the group is then $\lvert f_i \rvert^2=2P_j/(g_iP_{tot})$, by supposing power equipartition within each group (see Methods Sec \ref{subsec:subsec6_3}).

Fig.\ref{fig:Fig2}a shows that, in the linear propagation regime (i.e., for $E_{in}\leq 0.81$ nJ) most of the pulse energy is carried by the first 6 mode groups under the experimental coupling conditions, and mostly in the second group.
On the other hand, we can see in Fig.\ref{fig:Fig2}c that for $E_{in}=2.45$ nJ, the lower-order modes attract power from the HOMs as a consequence of the increasing IM-FWM. At this energy levels, a  strongly nonlinear quasi-soliton regime occurs.

At 4.10 nJ pulse energy (Fig.\ref{fig:Fig2}e), the fundamental mode has attracted half of the total power, and the propagation is approaching to a multimode soliton. Still, the soliton has not fully formed and does not show the characteristic Raman delay of a full soliton regime \cite{Zitelli2021a}; for this reason, we may neglect the presence of Raman scattering and other dissipative effects up to about this level of input pulse energy.

In the experiments, we also observed that for $E_{in}\ge 4.6$ nJ (not shown), a soliton is formed in the fundamental mode: for larger values of $E_{in}$, the soliton is further delayed in time and splits away from the remaining pulse carried by HOMs, because of the Raman soliton self-frequency shift \cite{Gordon1986,Renninger2013,Wright:15,Zitelli_9887813}. 


As shown in Figs.\ref{fig:Fig2}b, d, the output mode power distribution leads to a relatively broad and speckled output beam intensity profile at both low and intermediate values of $E_{in}$. Whereas approaching to the soliton regime, most of the energy is attracted by the fundamental mode; as a result, Fig.\ref{fig:Fig2}f shows that the beam brightness is substantially enhanced at its center: a bell-shaped beam is formed with a waist close to that of the fundamental mode, sitting on a background of HOM, a typical signature of beam self-cleaning \cite{krupa2017spatial}.

\begin{figure}[t!]
\includegraphics[width=0.95\textwidth]{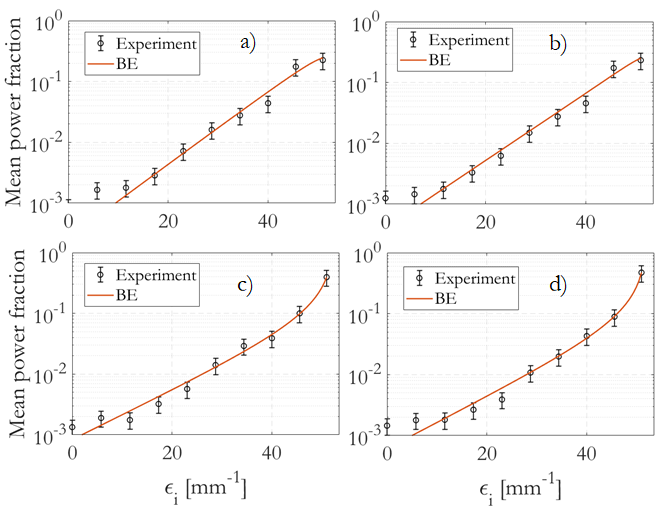}	\centering	
\caption{Average mode power fraction $\lvert f_i \rvert^2$ vs. modal eigenvalue $\epsilon_i$, for input pulse energies $E_{in}$ equal to a) 0.16 nJ,(b) 0.82 nJ, (c) 2.45 nJ, and d) 4.10 nJ, respectively. Corresponding analytic fits with the weighted BE distribution are shown by a solid orange curve.}
\label{fig:Fig3}
\end{figure}

Thanks to the accuracy offered by the used decomposition method, Fig.\ref{fig:Fig3} reports in log scale
the average output power fractions $\lvert f_i \rvert^2$ from Fig.\ref{fig:Fig2}, versus their respective mode eigenvalues $\epsilon_i$. In Fig.\ref{fig:Fig3}, the input pulse energy ranges from $E_{in}=0.2$ nJ, where mode mixing originates from RMC, up to  $E_{in}=4.1$ nJ, where IM-FWM is the dominant mechanism for transferring energy among nondegenerate modes \cite{krupa2017spatial}. 

Output modal distributions are fitted using Eq.~\ref{eq:BE}, taking care to respect the constraint for the normalized power $\gamma$. At low pulse energy (Fig.\ref{fig:Fig3}a), the weighted BE approximates to a straight line, in good agreement with predictions obtained by numerically solving the so-called power-flow equations \cite{Savovi2019PowerFI} \cite{Gloge:6774107} (see Sec \ref{subsec:sec3}). For pulse energy of 0.81 nJ (Fig.\ref{fig:Fig3}b), propagation is initially nonlinear, and later dominated by RMC. The weighted BE properly fits the experimental data up to the 9-th group order, for $\mu'=-58.2$ mm$^{-1}$, $T=8.71$ mm$^{-1}$ and $\gamma=3.60$.
In the figure, we observe that group 2 has larger energy fraction respect to the fitting equation. It will be shown later that a local condensation of energy into lower groups is obtained at this power level.

On the other hand, Figs.\ref{fig:Fig3}c, d show that, as soon as $E_{in} \ge 2.45$ nJ, the population of the fundamental mode grows larger, so that it preferentially acquires power from HOMs as a consequence of IM-FWM. In both cases, the weighted BE fits the experimental data up to the soliton regime, obtaining $\mu'=-53.1 (-52.2)$ mm$^{-1}$, $T=8.71 (10.14)$ mm$^{-1}$ and $\gamma=10.8 (18.0)$ for $E_{in}=2.45 (4.10)$ nJ, respectively; 
the BE distribution denotes a new equilibrium state that is induced by the strong nonlinearity, which is cumulated by the pulse over the entire fiber length. 
Note that, for $E_{in}=4.1$ nJ one has that $(\mu'+\epsilon_1)/T\simeq -0.13$,  which means that the RJ approximation to the BE law (i.e., $\lvert f_i \rvert ^2\propto -T/(\mu'+\epsilon_i)$) is appropriate around the fundamental mode with $i=1$. 



\begin{figure}[t!]
\includegraphics[width=0.95\textwidth]{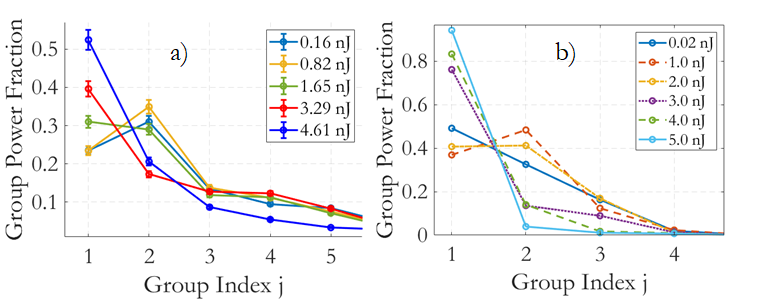}	\centering	
\caption{a) Experimental group power fraction $j \cdot \lvert f_j \rvert^2$ 
vs. group index $j$ for energies ranging between 0.16 nJ and 4.61 nJ. b) Simulated group power fraction, in similar conditions to the experiments.}
\label{fig:Fig7}
\end{figure}

Table \ref{tab1} shows the values of the optical temperature $T$, differential potential $\mu'$, factor $\gamma$ and state equation error $\epsilon_{SE}$ for the experiment of Fig.\ref{fig:Fig3}. For growing input energy $E_{in}$, temperature $T$ increases from 7.11 mm$^{-1}$ to 10.36 mm$^{-1}$; correspondingly, $\mu'$ reduces from -65.42 mm$^{-1}$ to -52.08 mm$^{-1}$; $\gamma$ scales proportionally to the input energy, and the error upon the state equation decreases from 7.9\% and 3.2\% in the linear regime down to 0.3\% in the nonlinear regime.

\begin{table}[h]
\caption{Thermodynamic parameters obtained fitting the data of Fig.\ref{fig:Fig3} using the weighted BE law.}\label{tab1}  \centering	
\begin{tabular}{@{}lllll@{}}
\toprule
$E_{in}$ [nJ] & $T$ [mm$^{-1}$] & $\mu'$ [mm$^{-1}$] & $\gamma$ & $\epsilon_{SE}$\\
\midrule
0.16 & 7.11	 & -65.42 & 0.72  & 0.0793 \\
0.41 & 7.76	 & -60.92 & 1.80  & 0.0318 \\
0.82 & 8.71	 & -58.19 & 3.60  & 0.0177 \\
1.23 & 9.32	 & -56.52 & 5.40  & 0.0122 \\
1.64 & 9.05	 & -54.74 & 7.20  & 0.0085 \\
2.06 & 9.53	 & -54.17 & 9.00  & 0.0069 \\
2.45 & 8.71	 & -53.10 & 10.80 & 0.0058 \\
3.29 & 9.75	 & -52.81 & 14.40 & 0.0041 \\
4.11 & 10.14 & -52.29 & 18.00 & 0.0035 \\
4.61 & 10.36 & -52.08 & 20.16 & 0.0035 \\
\botrule
\end{tabular}
\end{table}

A detailed information on how the relative power fraction for groups of nondegenerate modes, or $j \cdot \lvert f_j \rvert^2$, evolves as a function of group index $j$ for several $E_{in}$ is given in Fig.\ref{fig:Fig7}a.  
In the figure, it appears that in the intermediate nonlinear regime (0.82 nJ), group 2 locally attract energy at the expense of the HOMs; at 1.65 nJ and 3.29 nJ, local condensation into group 4 is observed.
When approaching to the soliton regime (4.61 nJ), the fundamental mode power grows up larger at the expenses of groups 2-4 and the HOMs, until it carries more than 50\% of the total output pulse energy.
For $E_{in}>4.6$ nJ, a Raman soliton or a train of solitons form, which can be spectrally isolated from the residual pulses; in this regime (not shown), experiments confirm that more than 90\% of the pulse energy is concentrated to the ground state.

Fig.\ref{fig:Fig7}b reports similar results from the numerical simulations explained in section \ref{subsec:sec4}. Here, it is evident the local condensation into group 2 at 1.0 nJ, and into group 3 at 3.0 nJ, followed by a global condensation into group 1 at 5.0 nJ.

In the supplementary Sec. \ref{sec:secA1}, a validation of the weigthed BE law is reported against independent experimental data from \cite{pourbeyram2022direct}.




\section{Simulations} \label{sec:sec3_0}

\subsection{Linear disorder} \label{subsec:sec3}

In the linear propagation regime, modal power exchanges originate from RMC. A well-know model to simulate linear coupling among modes in a GRIN fiber is provided by the power-flow equations  \cite{Savovi2019PowerFI} \cite{Gloge:6774107} (see Sec. \ref{subsec:subsec6_3}). The model describes a bi-directional flow of power, from group $j$ down to group $j-1$ and up to group $j+1$. After a number of consecutive integration steps, a cascading effect is produced, leading to power flow into non-adjacent groups. 
As discussed in Methods Sec. \ref{subsec:subsec6_3}, the mode coupling coefficients are not symmetrical in the two directions, resulting into a preferential transfer of power from HOMs into low-order modes of the fiber. 

\begin{figure}[h]
\includegraphics[width=0.95\textwidth]{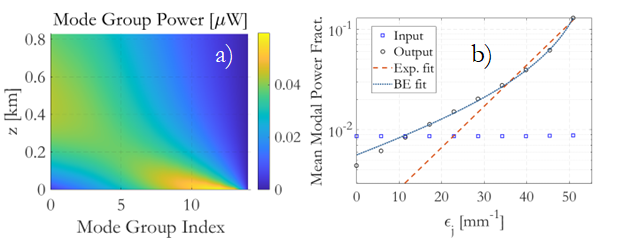}	\centering	
\caption{a) Power of the mode groups vs. distance, resulting from a power-flow simulation in the linear regime. b) Corresponding input and output mean modal power fractions vs. the modal eigenvalues.}
\label{fig:Fig6}
\end{figure}

An example of numerical solution of the power-flow equations is provided in Fig.\ref{fig:Fig6}. Here we simulated the propagation of the first 10 mode groups at 1400 nm, over 830 m of GRIN fiber. The linear loss is $\alpha_0=2.6$ dB/km, and the coupling coefficient is $D=0.043$ m$^{-1}$; we also included the presence of weak modal losses via the coefficient $A=2\times10^{-5}$ m$^{-1}$. The modal content at the fiber input is assumed to be equi-distributed; hence, the group power increases linearly with group index $j$. 

Fig.\ref{fig:Fig6}a shows the evolution of the power of the groups vs. distance: a net flow of power in the direction of the lower-order groups is observed. As a result, the output mean modal power fraction $\lvert f_i \rvert^2$ vs. the mode eigenvalue progressively increases (Fig.\ref{fig:Fig6}b). This distribution can be fitted well by the weighted BE law with $\mu'=-54.6$ mm$^{-1}$, $T=31.1$ mm$^{-1}$ and $\gamma=70.5$, which describes the RMC-induced steady-state. Figure also shows that a simple exponential function, of the type $\lvert f_i \rvert^2=a\exp{(-b\epsilon_i)}$ cannot properly fit the HOM groups for low losses. 

\subsection{Linear disorder and nonlinearity} \label{subsec:sec4}

In order to provide additional insight into the mode coupling dynamics leading to our experimental results, it is necessary to introduce a model which takes into account the presence of both linear and nonlinear mode coupling. To this end, we carried out extensive numerical simulations by using the coupled-mode GNLSEs \cite{Poletti2008a}, including wavelength-dependent linear and modal losses, and a new model for RMC which is derived from the power-flow equations of a GRIN fiber \cite{Savovi2019PowerFI}, as explained in the Method Sec.~(\ref{subsec:subsec6_4}). To save computation time, only $M=28$ modes were propagated over 100 m of GRIN fiber; we considered an input 250 fs pulse at 1400 nm, with the same coupling conditions as in the experiments. Modes 1 to 28 correspond to the Laguerre-Gauss modes $LG_{01}$, $LG_{11e}$, $LG_{11o}$, $LG_{21e}$, $LG_{21o}$,..., and $LG_{04}$, respectively. 
We set the RMC coupling coefficient $D=0.003$ m$^{-1}$, which introduces a significant amount of linear disorder over the considered distance. The RMC step was equal to $L_c=6$ mm, to ensure appropriate simulation accuracy. The fiber parameters (second and third-order chromatic dispersion, modal dispersion, Kerr and Raman nonlinearity, linear losses) correspond to typical values for an OM4 GRIN fiber at 1400 nm.

\begin{figure}[h]
\includegraphics[width=1\textwidth]{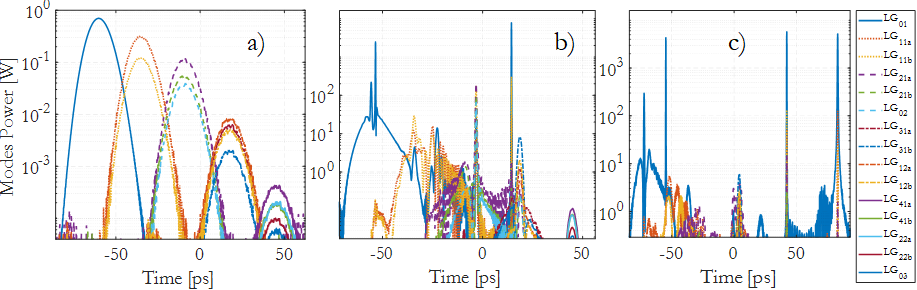}	\centering	
\caption{Simulated output modal power for $E_{in}$ equal to a) 0.02 nJ, b) 3.0 nJ and c) 5.0 nJ, respectively, with the same input coupling conditions as in the experiment.}
\label{fig:Fig8}
\end{figure}

Fig \ref{fig:Fig8}a shows the simulated temporal dependence of the output modal power in the linear regime ($E_{in}=0.02$ nJ). As can be seen, in each mode group the input pulses are broadened up to 13 ps by chromatic dispersion. Moreover, pulses in different mode groups separate in time by the action of inter-modal dispersion. 
On the other hand, Fig.\ref{fig:Fig8}b shows that in the nonlinear propagation regime ($E_{in}=3$ nJ), pulses within each mode group are significantly compressed in time by the combined action of SPM and anomalous dispersion; pulses do not experience Raman delay, meaning that a full soliton regime is not achieved yet. IM-FWM transfers power principally to the lower-order modes.


Finally, Fig.\ref{fig:Fig8}c shows that, in the soliton regime ($E_{in}=5$ nJ), ultrashort (170 fs duration) Raman delayed soliton pulses are formed; nearly 90\% of the power is attracted in the fundamental mode of all travelling pulses. A train of fundamental solitons results after long distance.

The resulting numerically simulated mode group power fraction $j \cdot \lvert f_j \rvert^2$ distribution at the GRIN fiber output is shown in Fig.\ref{fig:Fig7}b for the first 4 mode groups.
As can be seen,  in the linear propagation regime ($E_{in}=0.02$ nJ or 1.0 nJ), group power drops following a convex curve (in log scale). Whereas whenever Raman solitons are formed ($E_{in}=5.0$ nJ), the group power decreases following a concave curve: this is associated with a nonlinear irreversible energy condensation into the fundamental mode \cite{Zitelli2021}. At intermediate energy (2.0 to 3.0 nJ), one would expect a linear decreasing shape of the group power curve; to the contrary, local energy condensations are observed into groups 2 and 3. 

It must be said that the weighted BE is able to fit the output modal distributions (not shown) obtained from the simulations of Fig.\ref{fig:Fig7}b, denoting the achievement of steady states.

\section{Discussion and Conclusions}\label{sec:sec5}

The mean modal power fractions $\lvert f_i \rvert^2$ of Fig.\ref{fig:Fig3} show the achievement of steady states in the quasi-soliton regimes, because the weighted BE properly fits the distributions reached after long distances, as discussed in Sec. \ref{subsec:subsec6_1}.
Power-flow simulations in the linear regime, after long propagation distances, provide distributions properly fitted by the weighted BE as well, confirming that  Eq.~\ref{eq:BE} is a reliable method to analyze the achievement of steady states even at low power.

By representing the mean modal power fraction from the same modal group, against the input pulse energy (Fig.\ref{fig:Fig7}a), the fundamental mode (group 1) reaches the highest fraction in the soliton regime, where IM-FWM combined with RMC is responsible for power redistribution. For intermediate energy levels, between 0.4 and 2.0 nJ, groups 2 and 3 increase their fraction at the expense of the HOMs. A local attraction of energy to the lower-order groups is observed. 

The experimental findings of Fig.\ref{fig:Fig7}a are explained by observing results of the numerical simulations in Figs.\ref{fig:Fig8} and \ref{fig:Fig7}b. In all regimes (low power, quasi-soliton and soliton), RMC transfers power to modes within the same or of the neighbouring groups; in the linear regime, this process has the macroscopic effect to produce a BE distribution. 
However, in the quasi-soliton and soliton regimes, the Kerr nonlinearity and anomalous chromatic dispersion considerably shorten pulses; RMC diffuses power to the lower-order modes, a process which is boosted by the IM-FWM.


Globally, thanks to the interplay of RMC and IM-FWM, energy clusters are formed in the lower-order groups in the quasi-soliton regime (Figs.\ref{fig:Fig8}b, \ref{fig:Fig7} and \ref{fig:Fig1}b); a train of quasi-soliton pulses is produced, each composed by the modes of one group $j$, plus a fraction of modes belonging to the lower adjacent groups and of the fundamental mode; this results in the promotion of the first 3-4 modal groups at the output, and power clustering. In the framework of such interplay, RMC invalidates the achievement of a global condensation state as described by a RJ law \cite{PhysRevLett.129.063901}; however, the formation of steady states associated to a local condensation is still possible, a process which is characterized by a weighted BE modal distribution.

When increasing the pulse energy up to the soliton regime (Figs.\ref{fig:Fig8}c and \ref{fig:Fig1}c), global modal condensation to the ground state is observed in all splitted pulses; nearly 90\% of the power is attracted to the fundamental mode, as a consequence of the interplay of RMC and IM-FWM. Pulsewidth reduces to 170 fs, which is typical of propagated walk-off solitons \cite{Zitelli2021a}. A train of fundamental solitons is eventually produced by a peculiar fission mechanism mediated by modal dispersion; then solitons are affected by Raman soliton self-frequency shift \cite{Zitelli_9887813}.

In thermodynamic terms, experimental and numerical data confirm that the linear regime distribution describes a gas of energy packets, which evolves into a local condensed, or glassy state in the intermediate nonlinear regime, and then into a globally condensed solid state in the soliton regime \cite{Jung2019MeasuringTO}. All of the observed distributions correspond to steady states, being their modal distributions properly fitted by the weighted BE.

\section{Methods}\label{sec:sec6}

\subsection{Theory}\label{subsec:subsec6_1}

In the main text it was considered an optical multimode system including $Q$ groups of degenerate modes, distributed over $g_j/2$ modes and 2 polarizations, with $j=1, 2, .., Q$; hence, $g_j$ is the degeneracy over two polarizations; in the special case of a GRIN fiber, it is $g_j=2, 4, 6,.., 2Q$; the number of modes and polarizations is $2M=Q(Q+1)$. Following the procedure proposed in \cite{Wu2019}, the population $n_j$ of energy packets into the j-th group over 2 polarizations, leads to a total number of combinations among groups and polarizations 

\begin{equation}
W=\prod_{j=1}^{Q}\frac{(n_j+g_j-1)!}{(g_j-1)!n_j!}  .
\label{eq:Wcombination}
\end{equation}

System entropy is defined as $S=ln(W)$; by applying the Stirling approximation, valid for $n_j+g_j-1>>1$ we obtain

\begin{equation}
S=\sum_{j=1}^{Q}(n_j+g_j-1)\big[ln(n_j+g_j-1)-1\big]-(g_j-1)\big[ln(g_j-1)-1\big]-n_j\big[ln(n_j)-1\big]   ;
\label{eq:Entropy}
\end{equation}

Global system thermalization corresponds to the extremization of Eq.~\ref{eq:Entropy} using Lagrange multipliers for the total number of particles $N=\sum_{j}n_j$ and total normalized energy $E=\sum_{j}\beta_j n_j$, being $\beta_j$ the modal propagation constants supposed equal for the degenerate modes

\begin{equation}
\frac{\partial}{\partial n_l}\Big[S+ \sum_{j=1}^Q \Big( a' n_j+ b' \beta_j n_j \Big) \Big]=0      ,
\label{eq:EntropyDerivative}
\end{equation}

which provides, with no approximations

\begin{equation}
ln\Big(1+\frac{g_j-1}{n_j}\Big)+a'+b' \beta_j =0      ;
\label{eq:DerivativeRes1}
\end{equation}

Eq.~\ref{eq:DerivativeRes1} is valid for

\begin{equation}
\frac{n_j}{g_j-1}=\frac{1}{\exp{[-(a'+b' \beta_j)]}-1}    .
\label{eq:DerivativeRes2}
\end{equation}

We can choose $a'=\mu/(Tn_0)$ and $b'=1/(Tn_0)$, with $n_0$ a reference number of energy packets (for example the value at the lowest tested power) and $N=\gamma n_0$. $T$ (1/m) is an optical temperature and $\mu$ (1/m) a chemical potential. We these choices, Eq.~\ref{eq:DerivativeRes2} can be written as

\begin{equation}
n_j=\frac{(g_j-1)}{\exp{\Big(-\frac{\mu+\beta_j}{Tn_0}\Big)}-1}    .
\label{eq:DerivativeRes2b}
\end{equation}

An alternative development of Eq.~\ref{eq:DerivativeRes2} consists in replacing $a'$ and $b'$ with non-factorizable constants $a$ and $b$ defined as

\begin{equation}
-(a+b \beta_j)=\ln{\Big[\frac{1}{n_0}\exp{\Big(-\frac{\mu+\beta_j}{T}\Big)}-\frac{1}{n_0}+1\Big]} \simeq \frac{1}{n_0} \Big[\exp{\Big(-\frac{\mu+\beta_j}{T}\Big)}-1 \Big] .
\label{eq:DerivativeRes3}
\end{equation}

The approximation in Eq.~\ref{eq:DerivativeRes3} is valid for $\lvert a+b \beta_j \rvert << 1$, which is less than 10$^{-5}$ in the experiments. By replacing into Eq.~\ref{eq:DerivativeRes2} we obtain

\begin{equation}
n_j=\frac{n_0(g_j-1)}{\exp{\Big(-\frac{\mu+\beta_j}{T}\Big)}-1}    .
\label{eq:BEraw1}
\end{equation}

The particular choice of $a$, $b$, $a'$ and $b'$ conserves the system's energy and power; in fact it results, for $a'$, $b'$

\begin{equation}
\sum_{j=1}^Q a' n_j=-\sum_{j=1}^Q \frac{\mu(g_j-1)}{\mu+\beta_j}  ,
\label{eq:LagrangeEquality1}
\end{equation}

\begin{equation}
\sum_{j=1}^Q b' \beta_j n_j=-\sum_{j=1}^Q \frac{\beta_j (g_j-1)}{\mu+\beta_j}  ;
\label{eq:LagrangeEquality2}
\end{equation}

and 

\begin{equation}
\sum_{j=1}^Q a' n_j + b' \beta_j n_j=-\sum_{j=1}^Q (g_j-1)=Q-2M   .
\label{eq:LagrangeEquality12}
\end{equation}

For $a$, $b$ it also results

\begin{equation}
\sum_{j=1}^Q (a + b \beta_j) n_j=-\sum_{j=1}^Q (g_j-1)=Q-2M   ,
\label{eq:LagrangeEquality34}
\end{equation}
 
\noindent hence, the choices of $a$, $b$ or $a'$, $b'$ are equivalent in terms of power and energy conservation.


When passing from Eq.~\ref{eq:DerivativeRes2b} to Eq.~\ref{eq:BEraw1} by replacing the $a'$, $b'$ with $a$ and $b$,
the extremization of the entropy is performed using non-factorizable multipliers directly related to the modal group eigenvalues. The entropy extrema assume a local significance, because both $\mu$ and $T$ are now related to the particular set of $\beta_j$ and cannot be isolated. The total system's power and energy is conserved in the same way as for the global procedure of Eq.~\ref{eq:DerivativeRes2b}. The solution  Eq.~\ref{eq:BEraw1} is able to indicate the achievement of energy minima by the individual modal groups, such as locally condensed states; the equation is suitable to fit experimental modal distributions as a whole, regardless of the power fluctuations caused by local condensates.


We now pass to the $i$-th mode in the $j$-th group; in the case of a GRIN fiber, there are $2M=Q(Q+1)$ modes and polarizations, and $i=1, 2,..M$; it is $i=1$ for $j=1$, $i=2, 3$ for $j=2$, $i=4, 5, 6$ for $j=3$ and so on, and $g_i=[2, 4, 4, 6, 6, 6,.., 2Q]$. The mean modal power fraction over two polarizations is obtained as $\lvert f_i \rvert^2=2 n_j/(\gamma n_0 g_i)$, with $\gamma=N/n_0$; $\gamma > 1$ must scale with the experimental power. 

Introducing the differential eigenvalues $\epsilon_i=\beta_i-\beta_{j=Q}$, referred to the higher mode, and $\mu'=\mu+\beta_{j=Q}$, eq.~\ref{eq:BEraw1} provides the weighted Bose-Einstein (BE) modal distribution

\begin{equation}
\lvert f_i \rvert ^2=\frac{2(g_i-1)}{g_i\gamma}\frac{1}{\exp\big(-\frac{\mu'+\epsilon_i}{T}\big)-1}    .
\label{eq:BE1}
\end{equation}

The constraints for $\gamma$ are to scale with the input power, and to respect the conservation law $\sum_{i=1}^{M}\lvert f_i \rvert ^2=1$.

Let $P$ and $P_0$ be the optical powers corresponding to $N$ and $n_0$ energy packets (for an optical pulse, it is the peak power or the energy); hence, $\gamma=N/n_0=P/P_0$. The system's internal energy is $U=-\sum_{j}\beta_j n_j P/N$ (W/m); the power $P=\sum_{j} n_j P/N$ (W). From Eqs.~\ref{eq:LagrangeEquality12} and \ref{eq:LagrangeEquality34} we get

\begin{equation}
\sum_{j=1}^Q (a + b \beta_j) n_j=Q-2M=\frac{\mu N}{T n_0}+\frac{1}{T n_0}\Big(-\frac{UN}{P}\Big)  ,
\label{eq:StateEquation1}
\end{equation}

\noindent which provides the state equation

\begin{equation}
U-\mu P=(2M-Q)P_0T  .
\label{eq:StateEquation2}
\end{equation}

From the local extremization problem, providing Eq.~\ref{eq:BEraw1}, it is not possible to derive the well-known Rayleigh-Jeans (RJ) distribution \cite{Wu2019} \cite{podivilov2022thermalization} under reasonable approximations. However, from the global extremization solution, Eq.~\ref{eq:DerivativeRes2} with the definitions of $a'$ and $b'$, we obtain

\begin{equation}
\lvert f_i \rvert ^2=\frac{2(g_i-1)}{g_i\gamma n_0}\frac{1}{\exp\big(-\frac{\mu'+\epsilon_i}{T n_0}\big)-1}    ;
\label{eq:BEtrad}
\end{equation}

\noindent the presence of $N=\gamma n_0$ at the denominator of Eq.~\ref{eq:BEtrad} makes the equation suitable for fitting the experimental data only for $N < 10$; for this reason, it is unusable. The problem is removed under the assumption $\lvert \mu'+\epsilon_i \rvert<<\lvert T n0 \rvert$, which brings directly to the RJ distribution

\begin{equation}
\lvert f_i \rvert ^2=-\frac{2(g_i-1)}{g_i \gamma}\frac{T}{\mu'+\epsilon_i}    .
\label{eq:RJ}
\end{equation}

In terms of fractional power $\lvert c_i \rvert ^2=P\lvert f_i \rvert ^2$ in [W], by choosing $T'=P_0T$ (W/m), Eq.~\ref{eq:RJ} provides. 

\begin{equation}
\lvert c_i \rvert ^2=-\frac{2(g_i-1)}{g_i}\frac{T'}{\mu'+\epsilon_i} .
\label{eq:RJ1}
\end{equation}

The RJ law demonstrates to be a good choice for describing experiments characterized by global condensation states, such as the well-known self-cleaning experiments \cite{krupa2017spatial,mangini2022statistical,podivilov2022thermalization}, while the more general weighted BE, Eq.~\ref{eq:BE1}, is suitable to fit the experimental data also in cases where local condensed states are achieved.


\subsection{Experimental setup}\label{subsec:subsec6_2}

Optical pulses at 1400 nm wavelength and 250 fs pulsewidth are generated by an optical parametric amplifier (OPA) fed by a femtosecond Yb laser, at 100 kHz repetition rate, and using a narrow-band optical filter. The input beam is attenuated, linearly polarized and passed through a $\lambda /4$ waveplate, in order to generate a circular state of polarization. The Gaussian shaped beam is then injected into variable length spans of OM4 GRIN fiber (1 m, 830 m and 5 km), with a waist $w_0$ of approximately 13 $\mu$m. The corresponding self-imaging induced beam compression factor $C=2z_p/(\pi\beta_0 w_0^2)=0.305$, where $\beta_0=2\pi n_0/\lambda$, $z_p=0.55$ mm the self-imaging period, and $n_0=1.46$ is the core index \cite{Karlsson:92,Ahsan:19}; the effective beam waist for determining the nonlinear coefficient is $w_e=\sqrt{C}w_0= 7.27\mu$m \cite{Ahsan:18,Ahsan:19}, close to the fundamental mode waist. The circular state of polarization is used at the input in order to minimize power exchanges among polarizations.
The input beam is laterally shifted with respect to the fiber axis by 10 $\mu$m, in order to increase the contribution from higher-order modes. 

Linear losses were measured as $\alpha=6.0\text{x}10^{-4}$ m$^{-1}$. According to Gloge theory \cite{Gloge:72}, bending losses remain negligible up to the first 10 mode groups ($\alpha_j <= 8\text{x}10^{-10}$ dB/km for group $j=1, 2,..,10$, and $\alpha >> 30$  dB/km for group $j>=11$); in agreement with the theory, 10 modal groups could be observed at the output of both 830 m and 5 km fiber spans.

Modal dispersion is responsible for the time delay of the different groups; measured delay among groups is 206 ps over 830 m, and 850 ps over 5 km. Hence, mode groups are time-resolved at the output and easily measurable after 830 m, once the modes have interacted linearly and nonlinearly along the fiber.
Delay among groups was found to vary with distance as $\Delta t_j=1.02z^{0.79}$; RMC affecting the experiment is intermediate between weak (where modal delay scales with $z$) and strong regime (where it changes with $\sqrt{z}$) \cite{Marcuse1973LossesAI}.

At the fiber output, near-field is imaged on an InGaAs camera (Hamamatsu C12741-03); the beam is also directed to a real-time multiple octave spectrum analyzer with a spectral detection range of 1100-5000 nm (Fastlite Mozza). The output pulse instantaneous power is detected by a fast photodiode (Alphalas UPD-35-IR2-D) and a real-time oscilloscope (Teledyne Lecroy WavePro 804HD) with 30 ps overall response time. An intensity autocorrelator (APE pulseCheck 50) with femtosecond resolution is also used for the characterization of the input pulses. Input and output power are measured by a power meter with $\mu$W resolution.

Traditional 2D modal decomposition methods \cite{Flamm2012} are not suitable for the analysis of the output near-field after hundreds of meters of pulse propagation, because they do not account for: (i) the phase chirp which is induced by chromatic dispersion of pulses carried by the different modes; (ii) the phase delay among modes due to the modal dispersion; (iii) laser-induced phase noise; (iv) random phase differences among modes, introduced by the RMC.
In this work, we use the 3D modal reconstruction method proposed in \cite{Zitelli:23}; the mode group power is measured from the instantaneous power detected by a fast photodiode; comparison to the measured near-field is performed after 3D reconstruction. Such method has provided an accurate estimate of the modal distributions reached in long spans of GRIN, both in linear and nonlinear regime, up to the 10-th modal group (55 modes per polarization).

\subsection{Power-flow numerical model}\label{subsec:subsec6_3}

In the linear regime, RMC is properly modelled by the well-known power-flow diffusive equations. If $P_j$ is the power of the $j$-th mode group, the power exchange among adjacent modal groups is described by \cite{Savovi2019PowerFI}

\begin{multline}
P_j(z+L_c)=\big(\frac{DL_c}{\Delta m^2}-\frac{DL_c}{2m\Delta m}\Big)P_{j-1}(z)+ \\\Big(1-\frac{2DL_c}{\Delta m^2}-(\alpha_0+Am^2)L_c\Big)P_j(z)+
\big(\frac{DL_c}{\Delta m^2}+\frac{DL_c}{2m\Delta m}\Big)P_{j+1}(z)   ,
\label{eq:PowerFlow}
\end{multline}

\noindent with $m(j)=j-1$, $D$ and $\alpha_0$ (1/m) the coupling coefficient and linear losses, respectively, $A$ the modal loss coefficient, $\Delta m=1$ the modal step and $L_c$ the RMC integration step. In Eq.~\ref{eq:PowerFlow}, power flows in both directions, from group $j$ down to group $j-1$ and up to $j+1$; after consecutive integration steps, a cascading effect is produced, causing the coupling of non-adjacent groups. 

Mode coupling into groups is neglected, because it is so fast that statistical modal equipartition into groups can be assumed; it is then meaningful to calculate the mean modal content into groups as $\lvert f_i \rvert^2=2P_j/(g_iP_{tot})$.

\subsubsection{Modal power equipartition}

It is possible to demonstrate, using the alternative Gibb's definition of entropy \cite{doi:10.1119/1.1971557}, that modal equipartition does not generally apply among different groups. Given, at steady state, $p_i=<\lvert f_i \rvert^2>$ the modal occupation probability, and $\lambda_1, \lambda_2$ two Lagrange multipliers, extremization of the entropy, while power $P$ and energy $E$ are conserved, reads

\begin{equation}
\frac{\partial}{\partial p_k}\Big[-\sum_{i=1}^M p_i \log p_i +\lambda_1 \Big(\sum_{i=1}^M p_i-1 \Big) +\lambda_2 \Big(\sum_{i=1}^M p_i \beta_i-E/P \Big) \Big]=0      ;
\label{eq:EntropyDerivative2}
\end{equation}

\noindent from Eq.~\ref{eq:EntropyDerivative2}, if only the power is conserved ($\lambda_2=0$), we obtain  

\begin{equation}
\sum_{i=1}^M p_i =\sum_{i=1}^M \exp{(\lambda_1-1)}= M \exp{(\lambda_1-1)}=1  ,
\label{eq:Entropy2der1}
\end{equation}

\noindent which brings to the modal equipartition $p_i=1/M$. However, when considering also the conservation of the energy, Eq.~\ref{eq:EntropyDerivative2} provides

\begin{equation}
\sum_{i=1}^M p_i =\sum_{i=1}^M \exp{(\lambda_1+\lambda_2 \beta_i-1)}= 1  ,
\label{eq:Entropy2der2}
\end{equation}

\noindent which provides equipartition only for degenerate modes, with same $\beta_i$.



\subsection{GNLSE numerical model}\label{subsec:subsec6_4}

In the nonlinear (and linear) regime, numerical simulations use the coupled GNLSEs \cite{Poletti:08}, modified to include modal and wavelength-dependent losses, and linear random-mode coupling (RMC) for mode $p$

\begin{multline}
\frac{\partial A_p(z,t)}{\partial z} =i(\beta_0^{(p)}-\beta_0)A_p-(\beta_1^{(p)}-\beta_1) \frac{\partial A_p}{\partial t}+i\sum_{n=2}^{4} \frac{\beta_n^{(p)}}{n!}\big(i\frac{\partial}{\partial t}\big)^n A_p-\frac{\alpha_p(\lambda)}{2}A_p+ \\ 
+i\sum_m q_{mp}A_m+in_2k_0\sum_{l,m,n}S_{plmn} \big[(1-f_R)A_lA_mA_n^*+f_R A_l[h_R*(A_mA_n^*)]\big]  .
\label{eq:CoupledNLSE}
\end{multline}

In Eq.~\ref{eq:CoupledNLSE}, $\beta_n^{(p)}$ is the n-th order dispersion term (modal and chromatic) for mode $p$, $\alpha_p(\lambda)$ the modal and wavelength-dependent loss coefficient, $n_2$ (m$^2$/W) the nonlinear index coefficient multiplying the Kerr and Raman terms, $S_{plmn}$ is an overlap integral among modes, accounting for IM-FWM, and $q_{mp}$ is the linear RMC coupling coefficient, from mode $m$ to $p$, coming from the power-flow equations model \cite{Savovi2019PowerFI}

\begin{equation}
q_{mp}=\left\{
\begin{array}{cc}
    \big[\frac{DL_c}{p-1} \big( 1-\frac{1}{2(p-1)}\big)\big]^{1/2} & \text{from modes } m \text{ with }g_m=g_p-1\\ 
    \big[\frac{DL_c}{p+1}\big(1+\frac{1}{2(p-1)}\big)\big]^{1/2} & \text{from modes } m \text{ with }g_m=g_p+1  ,
\end{array}
\right.
\label{eq:RMCcoeff}
\end{equation}

\noindent being $g_p$ the degeneracy of modes, $p=1, 2,.., M$, $D$ (m$^{-1}$) the RMC coupling coefficient, and $L_c$ the RMC numerical step. Degenerate modes are not accounted for in Eq.~\ref{eq:RMCcoeff}, because their coupling is so fast that power equipartition can be assumed into groups.

In the main text, fiber parameters at $\lambda=1400$ nm are: $\beta_2=-11.8$ ps$^2$/km, $\beta_3=0.102$ ps$^3$/km for the fundamental mode, $\alpha=2.6$ dB/km, negligible modal loss, $n_2=2.7\text{x}10^{-20}$ m$^2$/W, $D=0.003$ (1/m), $L_c=6$ mm, $f_R=0.18$, $S_{plmn}$ calculated from modal overlap integrals.

---------------------

\backmatter



\bmhead{Acknowledgments}

The authors wish to thank C. Conti and G. D'Aguanno for valuable discussions about glass states, S. Savovic' for information regarding the power-flow model, M. Ferraro, D. Christodoulides, G. Pyrialakos and G. Steinmeyer for discussion regarding optical thermodynamics, and F. Wise and L. Wright for making freely available the open-source parallel numerical mode solver for the coupled-mode nonlinear Schrödinger equations \cite{Wright2018a}, and the test data used in supplementary Sec. \ref{sec:secA1}; a modified version of the software was used for the numerical simulations.

\section*{Declarations}


\begin{itemize}
\item Funding

European Union under the Italian National Recovery and Resilience Plan (NRRP), Rome Technopole Flagship Project 5.

HORIZON EUROPE European Research Council (101081871). 

NextGenerationEU, partnership on “Telecommunications of the Future” (PE00000001 - program “RESTART”).

\item Conflict of interest/Competing interests 

The authors declare no conflict of interest.

\item Ethics approval 
Not applicable
\item Consent to participate
Not applicable
\item Consent for publication
Not applicable
\item Availability of data and materials

Data available under request.

\item Code availability 
Not applicable
\end{itemize}

\noindent






\begin{appendices}

\section{Validation of the weighted BE law against independent experiments}\label{sec:secA1}

\begin{figure}[h]
\includegraphics[width=0.7\textwidth]{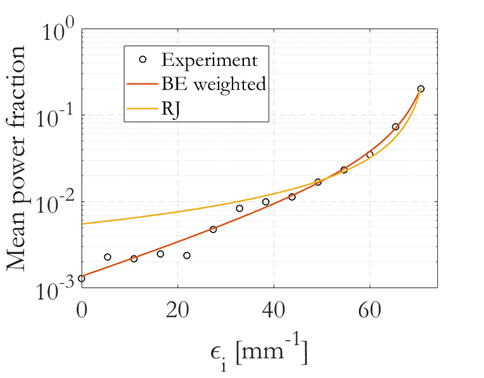}	\centering	
\caption{Experimental mean modal power fractions from Fig. 2b of \cite{pourbeyram2022direct}. RJ and weighted BE fits are reported.}
\label{fig:FigSupp1}
\end{figure}

In order to provide an independent validation of the weighted BE law introduced in Sec. \ref{sec:sec1}, we referred to the tests performed in \cite{pourbeyram2022direct} whose experimental data are published. Fig. 2b of that work is an experiment performed using 200 fs pulses at 1040 nm, propagating over 50 cm of GRIN fiber with 50 $\mu m$ diameter; 14 modal groups were measured with differential eigenvalues $\epsilon_i$ ranging between 0 and 70671 m$^{-1}$; thermalization was obtained in the experiment with 52 kW peak power. The experimental mean modal power fractions $\lvert f_i \rvert ^2$ are reported in Fig.\ref{fig:FigSupp1}; degenerate modes are supposed with same modal power fraction and eigenvalue. 

Numerical fits were performed using the weighted BE, Eq.~\ref{eq:BE}, and the RJ, Eq.~\ref{eq:RJ} with $\gamma=1$. The RJ fit provided $T=400$ m$^{-1}$ and $\mu'=-72600$ m$^{-1}$, equal to the values calculated in the original work; the RJ law appears following the distribution up to the 7-th modal group, but it does not follow correctly the distribution of the HOMs; the fit accuracy was $R^2=0.959$. The weighted BE fit obtained $T=23300$ m$^{-1}$, $\mu'=-71100$ m$^{-1}$ and $\gamma=70.1$; the curve appears following the distribution up to the 14-th group order, with accuracy $R^2=0.998$.

Fig. 3e in \cite{pourbeyram2022direct} is a second experiment performed using 200 fs pulses at 1040 nm, peak power 20 kW, propagating in 2.5 m of GRIN fiber with 62.5 $\mu m$ diameter. The measured modal groups were 28 with $\epsilon_i$ ranging between 0 and 176321 m$^{-1}$; the measured mean modal power fractions are reported in Fig.\ref{fig:FigSupp1}. The RJ fit provided in this case $T=292$ m$^{-1}$ and $\mu'=-178000$ m$^{-1}$ with accuracy $R^2=0.938$; in the original work, values were 300 and -178000 m$^{-1}$ respectively; the RJ is able to follow the experimental distribution up to the 8-th group order. The weighted BE fit obtained $T=45700$ m$^{-1}$, $\mu'=-179000$ m$^{-1}$ and $\gamma=166$; the fit provided an accuracy of $R^2=0.976$ and it was able to overlap to the experimental distribution up to the 17-th group order. 

A validation of the state equation Eq.~\ref{eq:StateEquation3} was performed calculating the error $\epsilon_{SE}$ from Eq.~\ref{eq:StateEquationError}; the results were 0.64 \% and 1.4 \% for the two experiments, respectively. 

\begin{figure}[h]
\includegraphics[width=0.7\textwidth]{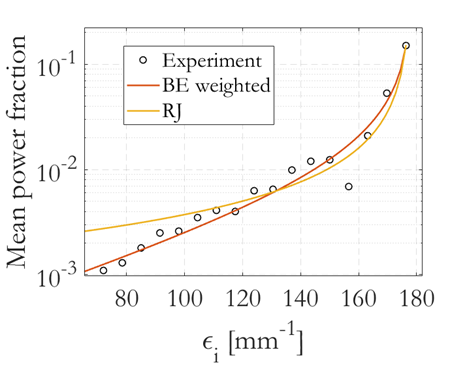}	\centering	
\caption{Experimental mean modal power fractions from Fig. 3e of \cite{pourbeyram2022direct}. RJ and weighted BE fits are reported.}
\label{fig:FigSupp2}
\end{figure}




\end{appendices}



\bibliographystyle{unsrt}
\bibliography{References.bib}



\end{document}